\documentclass{optica-article}

\journal{opticajournal} 

\articletype{Research Article}

\usepackage{lineno}
\usepackage{siunitx}
\usepackage{physics}
\usepackage{url}
\usepackage{hyperref}
\usepackage{threeparttable}

\begin{document}

\title{Metasurface-enabled non-orthogonal four-output polarization splitter for non-redundant full-Stokes imaging}

\author{Go~Soma,\authormark{1,*} Kento~Komatsu,\authormark{1} Chun~Ren,\authormark{1} Yoshiaki~Nakano,\authormark{1} and Takuo~Tanemura\authormark{1,$\dagger$}}

\address{\authormark{1}Department of Electrical Engineering and Information Systems, The University of Tokyo, 7-3-1 Hongo, Bunkyo-ku, Tokyo 113-8656, Japan}

\email{\authormark{*}go.soma@tlab.t.u-tokyo.ac.jp} 
\email{\authormark{$\dagger$}tanemura@ee.t.u-tokyo.ac.jp}

\begin{abstract*} 
Imaging polarimetry plays an essential role in various fields
since it imparts rich information that cannot be obtained through mere intensity and spectral measurements. 
To retrieve full Stokes parameters, at least four sensor pixels are required, each of which projects incident light to a different polarization state in the Stokes space.
Conventional full-Stokes division-of-focal-plane (DoFP) cameras realize this function by integrating angled polarizers and retarders on top of image sensors. 
Due to the inevitable absorption at the polarizers, however, the maximum efficiency of these schemes is limited to 50\% in theory.
Instead of polarizers, three sets of lossless polarization beam splitters can be used to achieve higher-efficiency polarimetry, however, at the cost of reduced spatial resolution due to the need for six redundant sensor pixels.
In this paper, we reveal, for the first time to our knowledge, that low-loss four-output polarization splitting (without filtering) is possible using a single-layer dielectric metasurface.
Although these four states
are not orthogonal to each other, our metasurface enables simultaneous sorting and focusing onto four sensor pixels with an efficiency exceeding 50\%, which is not feasible by a simple combination of space-optic components. 
The designed metasurface composed of silicon nanoposts is fabricated to experimentally demonstrate complete retrieval of full Stokes parameters at the near-infrared wavelength range from 1500 to 1600 nm with $-2.28$-dB efficiency. 
Finally, simple imaging polarimetry is demonstrated using a 3$\times$4 superpixel array.
\end{abstract*}

\section{Introduction}
Polarization is a fundamental property of light along with intensity and wavelength. The state of polarization (SOP) can be described using the four Stokes parameters, represented in a vector form as 
\begin{equation}
    \vec{S}=\mqty(S_0 \\ S_1 \\ S_2 \\ S_3)
    =\mqty(|\tilde{E}_x|^2+|\tilde{E}_y|^2 \\ |\tilde{E}_x|^2-|\tilde{E}_y|^2 \\ 2\Re(\tilde{E}_x^*\tilde{E}_y) \\ 2\Im(\tilde{E}_x^*\tilde{E}_y)),
    \label{eq:Stokes}
\end{equation}
where $\tilde{E}_x$ and $\tilde{E}_y$ are the complex electric-field amplitudes of $x$- and $y$-polarized components.
While conventional imaging systems, including our eyes, are insensitive to polarization, polarimetric imaging systems can capture SOPs over a scene of interest, providing a wealth of information, such as the shape and texture of surfaces \cite{Tyo2006-bi, Garcia2015-nf, Wolff1995-pa}. Owing to this unique functionality, polarimetric imaging has been used in a wide range of applications, including astronomy, remote sensing, biology, and computer vision \cite{Gastellu-Etchegorry2017-np, Hough2006-sc, Walraven1981-lc, Dubreuil2012-nh, De_Boer2017-wm, Garcia2018-ho, Powell2018-hf}.

Among several methods of polarimetric imaging, the division of focal plane (DoFP) scheme, which follows the same concept as standard color image sensors \cite{Bayer1976-sb}, is promising due to its low complexity, cost, and size \cite{Tyo2006-bi}. Indeed, compact DoFP polarization cameras have been developed \cite{Nordin1999-nr, Guo2000-me, Gruev2007-mq, Gruev2010-pi, Maruyama2018-yr} and commercialized \cite{SONY-PolCamera, Thorlabs-PolCamera} to detect linear polarization states of incident light.
As shown in Fig.~\ref{fig1}(a), these cameras integrate angled micropolarizers on top of image sensors to retrieve $S_1$ and $S_2$ components in Eq.~(\ref{eq:Stokes}).
By integrating an additional retarder layer as shown in Fig.~\ref{fig1}(b), an elliptical polarizer array can also be realized to enable full-Stokes detection, including $S_3$ component \cite{Myhre2012-lq, Hsu2014-eq, Hsu2015-hd}.
However, all of these schemes employ polarizers (i.e., polarization filters) that absorb the orthogonal SOPs, resulting in an inevitable 50\% transmission loss.

To realize full-Stokes imaging polarimetry with an efficiency exceeding the 50\% theoretical limit of conventional schemes, four polarizers can be replaced with three sets of ideally lossless polarization beam splitters (PBSs), each of which projects incident light to orthogonal SOPs along the $S_1$, $S_2$, and $S_3$ axes in the Stokes space \cite{Tanemura2020-nm}.
Such a six-output PBS-based configuration has been demonstrated compactly by using optical metasurfaces \cite{Arbabi2018-vz, Ren2022-bi, Huang2023-nl, Soma2023-fx}.
A metasurface consists of a two-dimensional array of scatterers (namely, meta-atoms) that can locally manipulate both the phase and polarization of incident light \cite{Kamali2018-jb, Chen2016-ja, Yu2014-we}. 
By judiciously designing the geometry of each meta-atom, various functionalities can be realized, including focusing
\cite{Lin2014-fb, Arbabi2015-tw, Khorasaninejad2016-eh, Khorasaninejad2017-sb, Arbabi2022-jk}, 
holographic imaging \cite{Ni2013-be, Zheng2015-xm, Wang2016-aj, Jiang2019-vn, Bao2022-mz},
polarization manipulation \cite{Arbabi2015-rv, Balthasar_Mueller2017-ex, Rubin2021-mz, Dorrah2021-yb, Hada2024-wh}, polarimetry
\cite{Arbabi2018-vz, Rubin2019-tb, Ren2022-bi, Huang2023-nl, Pors2015-il, Balthasar_Mueller2016-la, Wei2017-cx, Yang2018-yh, Zhang2019-zh, Li2020-ia, Miyata2020-oe, Soma2023-fx, Fan2023-qi}, and coherent receivers \cite{Komatsu2024-JLT, Komatsu2024-ofc}. 
By employing dielectric metasurfaces either in segmented \cite{Arbabi2018-vz, Ren2022-bi} or interleaved \cite{Huang2023-nl, Soma2023-fx} configurations, input light can be split at three polarization bases and focused onto six different sensor pixels.
Since no optical power is lost, this method can, in principle, break the 50\% theoretical efficiency limit of conventional approaches that use polarizers.
On the other hand, six redundant sensor pixels are required to retrieve the full Stokes parameters, resulting in reduced spatial resolution. 
In addition, rectangular \cite{Arbabi2018-vz}, hexagonal \cite{Ren2022-bi}, or sparsely spaced \cite{Huang2023-nl} superpixels are undesirable when implementing these schemes to practical polarization cameras. 

In this paper, we demonstrate a single-layer metasurface-based polarization splitter for full-Stokes DoFP imaging polarimetry using only four sensor pixels, which is the minimum redundant number of pixels to detect full Stokes parameters in theory \cite{Tyo2006-bi}.
Unlike conventional full-Stokes DoFP polarimetry using four elliptical polarizers [Fig.~1(b)] \cite{Hsu2014-eq, Hsu2015-hd, Myhre2012-lq}, our metasurface [Fig.~1(c)] is designed to serve as an ideally lossless four-output polarization splitter (without filtering) that projects input light onto four polarization states, corresponding to a regular tetrahedron inscribed in the Poincaré sphere, and focuses them to four different pixels at the focal plane.
Since these four polarization states are not orthogonal to each other, it is a nontrivial question whether such polarization splitting and focusing are possible without incurring loss. 
Indeed, a simple combination of traditional space-optic components cannot achieve such capability.
Nevertheless, we reveal, for the first time to our knowledge, both numerically and experimentally that a high efficiency exceeding the 50\% theoretical limit of conventional four-pixel polarimeters can be achieved through a proper metasurface design based on the matrix polar decomposition.
With the fabricated metasurface on a silicon-on-quartz (SOQ) substrate, we achieve successful retrieval of input Stokes parameters at a broad wavelength range from 1500 to 1600 nm. Applicability to imaging polarimetry is also demonstrated using a $3\times4$ superpixel array. 

We should note that similar concepts have been demonstrated using a diffractive metasurface grating \cite{Rubin2019-tb} and volumetric metaoptics operating at mid-infrared wavelength \cite{Roberts2023-qa}.
The former work relies on the projection of four polarization states to different diffraction orders of a metasurface grating to construct a division-of-aperture (instead of DoFP) camera; as a result, it suffers from inherent chromatic dispersion 
and requires a reimaging optics \cite{Tyo2006-bi} that complicates the system.
The latter work, on the other hand, employs a rather complicated three-dimensional metamaterial, which would be challenging to fabricate at a shorter wavelength and to be integrated compactly with image sensors. Our polarimeter, in contrast, employs a single-layer metasurface that can easily be integrated on an image sensor to realize a simple and compact DoFP full-Stokes camera with minimum redundancy.

\section{Device concept}
\begin{figure}[tb]
\centering\includegraphics[width=\textwidth]{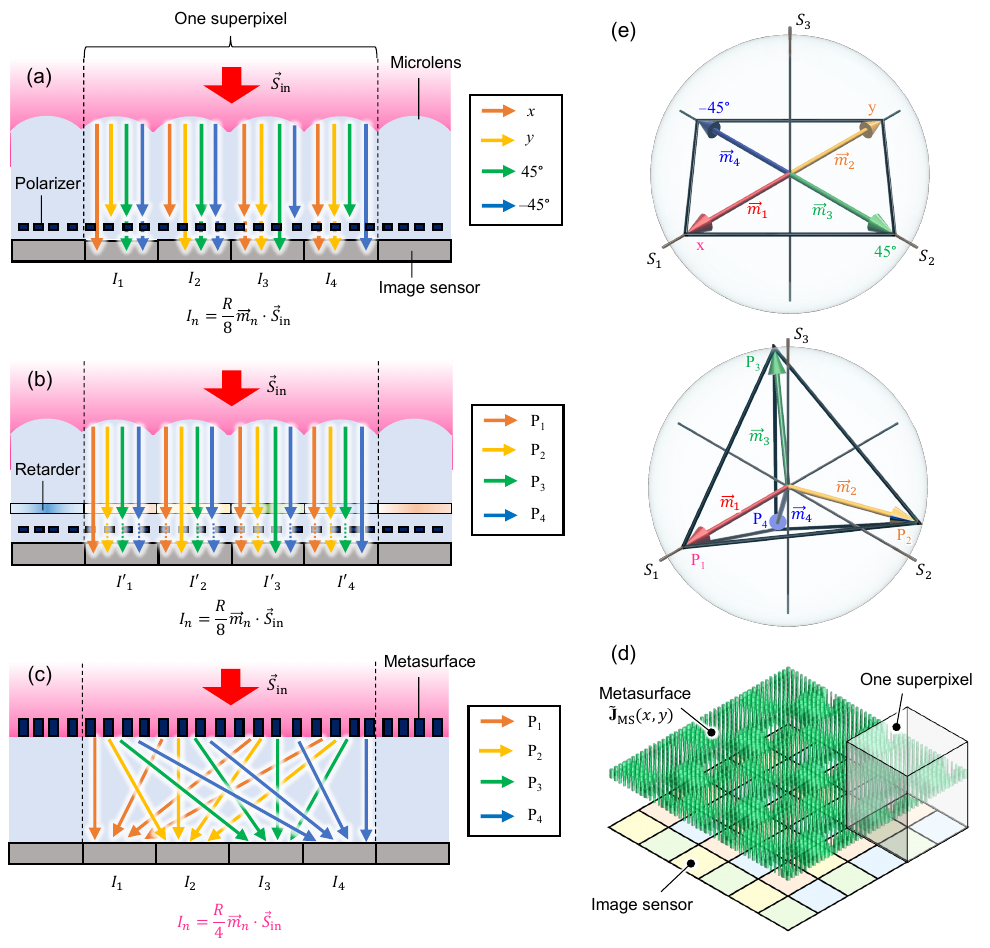}
\caption{Comparative descriptions of conventional and proposed polarimeters. (a-c) Schematic of polarimeters using (a) linear polarization filters, (b) elliptical polarization filters, and (c) our four-output polarization-sorting and focusing metasurface with zero inherent loss. (d) DoFP polarization camera composed of a 2D array of our metasurface-based polarimeters. (e) Arrangements of the four polarization states, P$_n$, on the Poincaré sphere: $(\vec{m}_{1}, \vec{m}_{2}, \vec{m}_{3}, \vec{m}_{4})$. The four Stokes vectors in our polarimeter construct a regular tetrahedron to achieve maximum sensitivity.}
\label{fig1}
\end{figure}

The schematic of our proposed full-Stokes polarimeter using a metasurface is shown in Fig.~\ref{fig1}(c). A DoFP polarization camera can be realized by arranging this metasurface in a two-dimensional (2D) array to form a square superpixel, as shown in Fig.~\ref{fig1}(d). 

When input light with a Stokes vector, represented as $\vec{S}_\mathrm{in}$, transmits through the metasurface, it is sorted by the polarization states, P$_1$, P$_2$, P$_3$, and P$_4$, and focused onto four different pixels of an image sensor. 
Here, we select P$_n$ to be the four vertices of a regular tetrahedron inscribed in the Poincaré sphere as shown in Fig.~\ref{fig1}(e) to maximize the sensitivity \cite{Tyo2006-bi, Sabatke2000-vb, Tanemura2020-nm}.
The full Stokes vector of the input light can then be calculated using the four signals, $\vec{I}=(I_1, I_2, I_3, I_4)^t$, detected at the sensor pixels.
Each signal $I_n$ is proportional to the intensity of the incident light projected to the corresponding polarization state P$_n$.
We define $\vec{m}_n=(1,m_{n1},m_{n2},m_{n3})^t$ as the normalized Stokes vectors describing P$_n$. Then, $I_n$ can be written as
\begin{equation}
    I_n=\frac{\eta R}{4}(\vec{m}_n\cdot\vec{S}_\mathrm{in}),
    \label{eq:I_n}
\end{equation}
where $R$ is the responsivity of the sensor and $\eta$ is a dimensionless proportional coefficient.
Eq.~(\ref{eq:I_n}) for all $n$ is expressed as
\begin{equation}
    \vec{I}=\frac{\eta R}{4}\ \mqty(
1 & m_{11} & m_{12} & m_{13} \\
1 & m_{21} & m_{22} & m_{23} \\
1 & m_{31} & m_{32} & m_{33} \\
1 & m_{41} & m_{42} & m_{43}
)\ \vec{S}_\mathrm{in}
\equiv \vb{A} \vec{S}_\mathrm{in},
\label{eq:I}
\end{equation}
where $\vb{A}$ is the $4\times4$ instrument matrix. From Eq.~(\ref{eq:I}), we can retrieve the input Stokes vector $\vec{S}_\mathrm{in}$ by simply multiplying the inverse matrix $\vb{B}=\vb{A}^{-1}$ to the detected signals $\vec{I}$. 

Since P$_n$ are set symmetrically in the Stokes space as shown in Fig.~\ref{fig1}(e), we have $\sum_{n=1}^{4} \vec{m}_n = (4,0,0,0)^t$. 
Therefore, the total detected signal at the four sensor pixels is written as
\begin{equation}
    I_\mathrm{total} \equiv \sum_{n=1}^{4} I_n=\eta R S_\mathrm{in0},
\end{equation}
where $S_\mathrm{in0}$ is the $S_0$ component of $\vec{S}_\mathrm{in}$ that denotes the input optical power.
Thus, $\eta$ actually represents the total transmission efficiency through the metasurface.
In the conventional DoFP cameras using polarization filters as shown in Fig.~\ref{fig1}(a) and \ref{fig1}(b), we can derive $\eta=1/2$ \cite{Tanemura2020-nm}.
This implies that we inevitably lose 50\% of the total incident optical power.
In contrast, in our polarimeter shown in Fig.~\ref{fig1}(c), we seek a metasurface design that is ideally lossless so that $\eta=1$.


\section{Metasurface design and numerical verification}

\begin{figure}[tb]
\centering\includegraphics{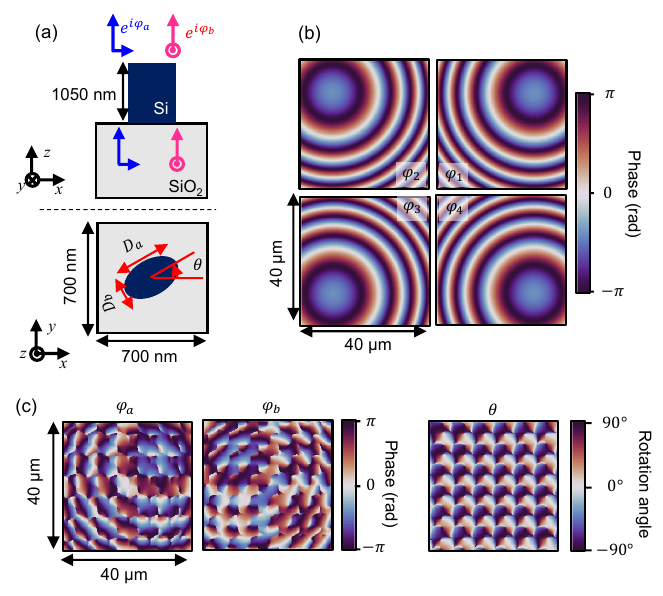}
\caption{Schematic and design of meta-atoms. (a) Si elliptical nanopost on a quartz substrate used as a meta-atom. (b) Required phase profiles $\varphi_n(x,y)$ for polarization states P$_n$ ($n=1,2,3,4$). (c) Spatial distributions of the required phase shifts $(\varphi_a(x,y),\varphi_b(x,y))$ and the orientation angle $\theta$ of meta-atoms.}
\label{fig2}
\end{figure}

As shown in Fig.~\ref{fig2}(a), we employ Si elliptical nanoposts on a quartz substrate as meta-atoms. Such meta-atom exhibits various degrees of birefringence depending on its geometry and can locally induce arbitrary phase shifts for two orthogonal linear polarization components of incident light. Thus, operation of each meta-atom at position $(x, y)$ can be represented using a $2\times2$ Jones matrix as \cite{Balthasar_Mueller2017-ex}
\begin{equation}
    \tilde{\vb{J}}_\mathrm{MS}(x, y)=\vb{R}(\theta(x, y))\mqty(
e^{i \varphi_a(x, y)} & 0 \\
0 & e^{i \varphi_b(x, y)}
) \vb{R}(-\theta(x, y)),
\label{eq:J_MS}
\end{equation}
where $\varphi_a(x, y)$ and $\varphi_b(x, y)$ are the phase shifts induced to the linearly polarized components along the fast and slow axes of the Si nanoposts, $\theta$ is its orientation angle, and $\vb{R}$ is a rotation matrix. 
We can show that any symmetric and unitary $2\times2$ matrix can be written in the form of Eq.~(\ref{eq:J_MS}) \cite{Rubin2021-mz}, which can physically be implemented by appropriately selecting the three parameters ($\varphi_a,\varphi_b,\theta$) of each meta-atom. 

Next, we consider the desired Jones matrix $\tilde{\vb{J}}_\mathrm{d}(x, y)$ to realize ideally lossless polarization-splitting function as explained in Section 2. The operation to select only P$_n$ polarization component of the incident light and focus it to a corresponding sensor pixel is expressed as
\begin{equation}
    \tilde{\vb{J}}_n(x, y)=e^{i\phi_n(x,y)}\tilde{\vb*{u}}'_{n}\tilde{\vb*{u}}_{n}^\dagger,
    \label{eq:J_n}
\end{equation}
where $\dagger$ represents the Hermitian adjoint. $\tilde{\vb*{u}}_{n}$ is the Jones vector of the projecting polarization state P$_n$, and $\tilde{\vb*{u}}'_{n}$ is the output Jones vector.
To physically realize $\tilde{\vb{J}}_n$ with a meta-atom as represented by Eq.~(\ref{eq:J_MS}), it needs to be a symmetric matrix.
For this condition to be satisfied,  $\tilde{\vb*{u}}'_{n}$ must be $\tilde{\vb*{u}}_{n}^*$. 
$\phi_n(x,y)$ indicates the required phase profile to focus the light to the center position $(x_n,y_n)$ of the corresponding sensor pixel and can be written as \cite{Khorasaninejad2016-eh}
\begin{equation}
    \phi_n(x, y)=-\frac{2 \pi}{\lambda}\qty(\sqrt{\qty(x-x_n)^2+\qty(y-y_n)^2+f^2}-f),
\end{equation}
where $\lambda$ is the wavelength and $f$ is the focal length. Figure~\ref{fig2}(b) shows the required phase profiles for the four polarization components. Here, we assume $\lambda =$ 1.55 \textmu m, $f=$ 80 \textmu m, and the size of the entire superpixel to be $40\times40$ \textmu m$^2$. 

Then, the desired Jones matrix of an ideally lossless four-output polarization splitter can be mathematically expressed as
\begin{equation}
    \tilde{\vb{J}}_\mathrm{d}(x, y)=\frac{1}{\sqrt{2}}\sum_{n=1}^{4} \tilde{\vb{J}}_n(x, y).
    \label{eq:J_d}
\end{equation}
We should note that the coefficient $1/\sqrt{2}$ describes the transmittance of the electric field amplitude for each polarization component and is derived as follows:
In Eq.~(\ref{eq:I_n}), $\vec{m}_n\cdot\vec{S}_\mathrm{in}=2S_\mathrm{in0}$ when the polarization state of the input light is P$_n$, i.e. $\vec{S}_\mathrm{in}=S_\mathrm{in0}\vec{m}_n$. In such a case, $I_n = \eta R S_\mathrm{in0}/2$, implying that the optical transmission for each polarization component is $\eta/2$. Therefore, in an ideal case of $\eta=1$ without the total loss, the transmittance of the electric field amplitude for each polarization component needs to be $1/\sqrt{2}$.
In contrast, for the case of conventional polarization-filter-based polarimeters, this coefficient is $1/2$, resulting in an inherent 3-dB overall loss.

Since $\tilde{\vb{J}}_\mathrm{n}$ in Eq.~(\ref{eq:J_n}) is chosen to be symmetric, $\tilde{\vb{J}}_\mathrm{d}$ described by Eq.~(\ref{eq:J_d}) is also symmetric. 
On the other hand, $\tilde{\vb{J}}_\mathrm{d}$ is not necessarily unitary, so it cannot generally be implemented by a single-layer metasurface described by Eq.~(\ref{eq:J_MS}). 
Thus, to extract the unitary matrix component from $\tilde{\vb{J}}_\mathrm{d}$, we perform the matrix polar decomposition \cite{Rubin2021-mz} as follows: $\tilde{\vb{J}}_\mathrm{d}$ can generally be decomposed as $\tilde{\vb{J}}_\mathrm{d}=\tilde{\vb{P}}\tilde{\vb{U}}$, where $\tilde{\vb{P}}$ is a positive semi-definite symmetric matrix (physically representing a partial polarizer) and $\tilde{\vb{U}}$ is a unitary matrix (representing a waveplate function).
Since $\tilde{\vb{J}}_\mathrm{d}$ is symmetric, $\tilde{\vb{U}}$ is also symmetric, implying that $\tilde{\vb{U}}$ can physically be implemented by a single-layer metasurface $\tilde{\vb{J}}_\mathrm{MS}$.
We, therefore, approximate $\tilde{\vb{J}}_\mathrm{d}$ by $\tilde{\vb{U}}$ ($=\tilde{\vb{J}}_\mathrm{MS}$) by simpliy discarding $\tilde{\vb{P}}$.
Such approximation is equivalent to the phase-matching approach for a scalar field, where the amplitude variation is neglected to design single-layered phase masks and metasurfaces \cite{Arbabi2015-tw}. 
From the above calculations, the required parameters of each meta-atom ($\varphi_a, \varphi_b, \theta$) are determined. Figure~\ref{fig2}(c) shows the calculated spatial distribution of these three parameters.

\begin{figure}[tb!]
\centering\includegraphics{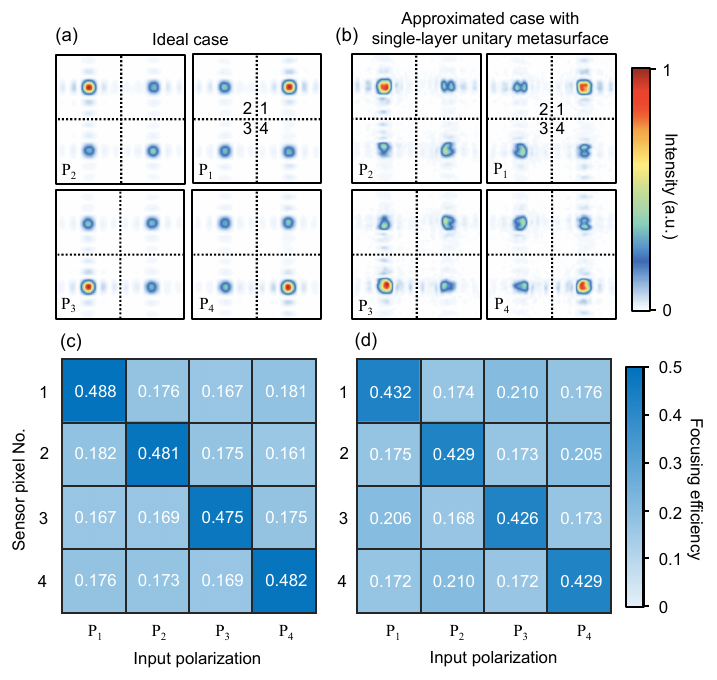}
\caption{Simulated results of designed metasurfaces. (a, b) Intensity distributions at the focal plane for the four different input SOPs, P$_n$ ($n=1,2,3,4$). (c, d) Focusing efficiencies to all pixels for each input SOP. (a) and (c) assume the ideal case with $\tilde{\vb{J}}_\mathrm{d}(x,y)$, whereas (b) and (d) assume the approximated case using a single-layer metasurface with $\tilde{\vb{J}}_\mathrm{MS}(x,y)$.}
\label{fig3}
\end{figure}

Numerical simulation based on the angular spectrum method (ASM) \cite{Goodman2017-mb} is used to validate our approach. Here, we assume the paraxial approximation and ignore the $z$ component of the electromagnetic fields. When a plane wave with a spatially uniform $2\times1$ Jones vector $\tilde{\vb*{j}}_\mathrm{in}$ is incident on the metasurface, the Jones-vector profile $\tilde{\vb*{j}}_\mathrm{f}(x, y)$ at the focal plane is given by
\begin{equation}
\begin{split}
    \tilde{\vb*{j}}_\mathrm{f}(x, y) 
    &=\mathcal{F}^{-1}\qty[ \tilde{H}_\mathrm{ASM}(k_x, k_y)\mathcal{F}\vb[\tilde{\vb{J}}_\mathrm{MS}(x,y)]]\tilde{\vb*{j}}_\mathrm{in},
\end{split}
\end{equation}
where $\mathcal{F}$ and $\mathcal{F}^{-1}$ represent the Fourier transform and the inverse Fourier transform, respectively. $\tilde{H}_\mathrm{ASM}(k_x, k_y)=\exp [i\sqrt{(2\pi/\lambda)^2-k_x^2-k_y^2}\ f]$ is a free-space propagator function from the metasurface to the focal plane.
By first calculating the Jones matrix distribution $\tilde{\vb{J}}_\mathrm{f}(x, y)\equiv\mathcal{F}^{-1}\qty[ \tilde{H}_\mathrm{ASM}\mathcal{F}\vb[\tilde{\vb{J}}_\mathrm{MS}]]$, the field distributions at the focal plane for arbitrary input polarization is obtained as $\tilde{\vb*{j}}_\mathrm{f}(x, y)=\tilde{\vb{J}}_\mathrm{f}(x, y)\tilde{\vb*{j}}_\mathrm{in}$.

We numerically calculate the intensity profiles at the focal plane for the input SOPs of P$_1$, P$_2$, P$_3$, and P$_4$ assuming two cases: the ideal case with the Jones matrix represented as $\tilde{\vb{J}}_\mathrm{d}$ and the physically feasible case approximated by $\tilde{\vb{J}}_\mathrm{MS}$. Figures~\ref{fig3}(a) and \ref{fig3}(b) show the simulated intensity distributions. We can confirm that the light is focused onto well-defined spots depending on the input SOP in both cases. From Fig.~\ref{fig3}(a) and \ref{fig3}(b), the focusing efficiencies to the four sensor pixels are calculated and shown in Fig.~\ref{fig3}(c) and \ref{fig3}(d). 
While we can observe some distortion in the intensity distribution and increased crosstalk to adjacent pixels for the case of approximated $\tilde{\vb{J}}_\mathrm{MS}$, the total transmission is higher than 98\% in both cases, exceeding the 50\% theoretical efficiency limit of the conventional polarizer-based polarimeters. 
For more quantitative evaluation, we calculate the sensitivity penalty $L$ for both cases (the detailed definition and derivation are given in Appendix~\ref{sec:L}).
The penalty is derived to be 0.31 and 1.35 dB for the ideal case with $\tilde{\vb{J}}_\mathrm{d}$ and the approximated case of using unitary meta-atoms with $\tilde{\vb{J}}_\mathrm{MS}$, respectively.
Once again, the penalty is less than the conventional theoretical limit of 3 dB in both cases.

\section{Fabrication of metasurface and experimental results}
\begin{figure}[tb]
\centering\includegraphics{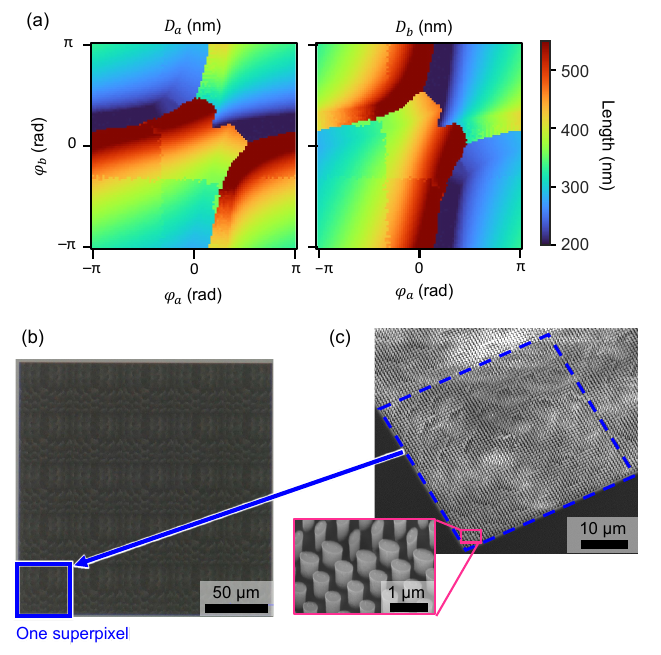}
\caption{Fabricated metasurface. (a) Lookup table used to determine the geometry of each meta-atom, showing required ($D_a$, $D_b$) to obtain given phase shifts $(\varphi_a,\varphi_b)$. 
(b) Optical microscope image and (c) scanning electron microscope (SEM) images of the fabricated metasurface.}
\label{fig4}
\end{figure}
We employed 1050-nm-high elliptical Si nanoposts with dimensions $(D_a, D_b)$, which were arranged on a square lattice with a lattice constant of 700 nm as shown in Fig.~\ref{fig2}(a). 
To determine the geometry of each meta-atom, we first calculated the transmission characteristics of a periodic uniform meta-atom array 
by a rigorous coupled wave analysis (RCWA) method \cite{Liu2012-kd}. 
From these results, we derived the look-up tables in Fig.~\ref{fig4}(a) that show the required $(D_a, D_b)$ to induce given phase shifts ($\varphi_a, \varphi_b$) (see Section~S1 in Supplement~1 for the details of calculation). We then derived $(D_a, D_b, \theta)$ of each meta-atom to implement desired $\tilde{\vb{J}}_\mathrm{MS}$.

The designed metasurface was fabricated on an SOQ substrate. First, the nanopost patterns were defined by electron-beam lithography with ZEP520A resist. Then, the patterns were transferred to the Si layer by the Bosch process based on reactive-ion etching (RIE) using SF$_6$ and C$_4$F$_8$, followed by the O$_2$ plasma ashing process. In this work, we fabricated a $5\times5$ superpixel array. Figures~\ref{fig4}(b) and \ref{fig4}(c) show an optical microscope image and scanning electron microscope (SEM) images of the fabricated metasurface. 
It was characterized by measuring the intensity distribution at the focal plane under a plane-wave illumination at 1550 nm wavelength. The incident polarization state was adjusted by a half-wave plate (HWP) and a quarter-wave plate (QWP) (see Section~S2 in Supplement~1 for the details of the optical setup).

\begin{figure}[tb]
\centering\includegraphics{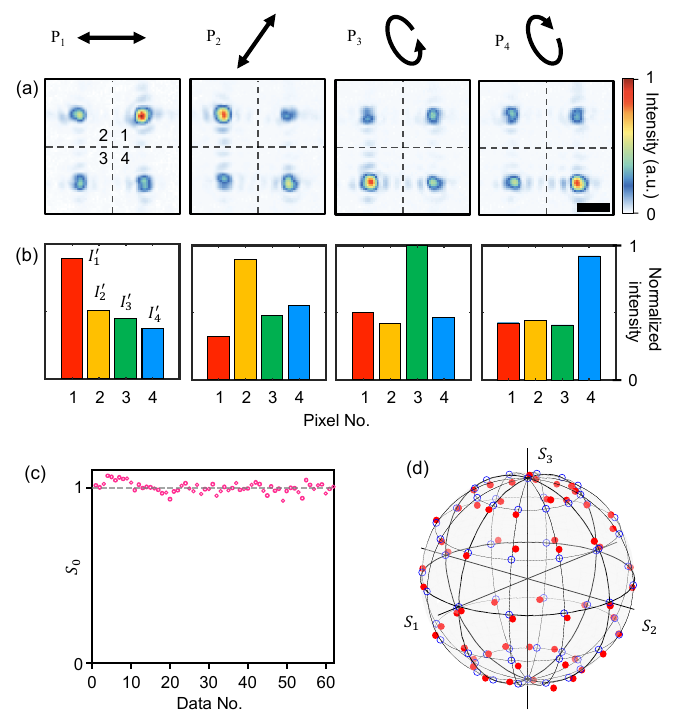}
\caption{Measured results of one superpixel. (a) Intensity distributions at the focal plane for the four different input SOPs, P$_n$ ($n=1,2,3,4$). (b) Integrated intensities at four pixels for each input SOP. (c,d) Retrieved $S_0$ and Stokes vector $(S_1, S_2, S_3)^t$ on the Poincaré sphere. Red: retrieved. Blue: actual.}
\label{fig5}
\end{figure}

Figure~\ref{fig5}(a) shows the measured intensity distribution of one superpixel at the focal plane for four polarization inputs.
The total light intensity at each sensor pixel 
is derived by integrating the measured intensity over the pixel area and plotted in Fig.~\ref{fig5}(b). 
We can confirm that the incident light is focused to the correct position for each input polarization. 
To minimize the effect of fabrication errors, we derive the actual retrieval matrix $\vb{B}'$, which deviated slightly from the designed $\vb{B} = \vb{A}^{-1}$ given in Eq.~(\ref{eq:I}) (detailed procedure to derive $\vb{B}'$ is given in Section~S3 in Supplement~1).
Figures~\ref{fig5}(c) and \ref{fig5}(d) show the retrieved full Stokes parameters using $\vb{B}'$ when we vary the input SOP over the entire Poincaré sphere. The average 
measurement error 
$\ev{|\Delta\vec{S}_\mathrm{ret}|}$ is as small as 0.078.
The measured transmittance $(T_x+T_y)/2$ is 0.858 in average, where $T_x$ and $T_y$ are the transmittance for the $x$- and $y$-polarized input, respectively. 
Using $\vb{B}'$, the sensitivity penalty $L$ is calculated to be 2.28 dB, which is below the 3-dB theoretical limit of polarization-filtering-based polarimetry.
The sensitivity degradation compared to the simulation is attributed to the reflection at the metasurface and residual fabrication errors, which could be mitigated by adopting a more advanced method to design geometries of each meta-atom \cite{Molesky2018-uw} and improving the fabrication process. 

\begin{figure}[tb]
\centering\includegraphics{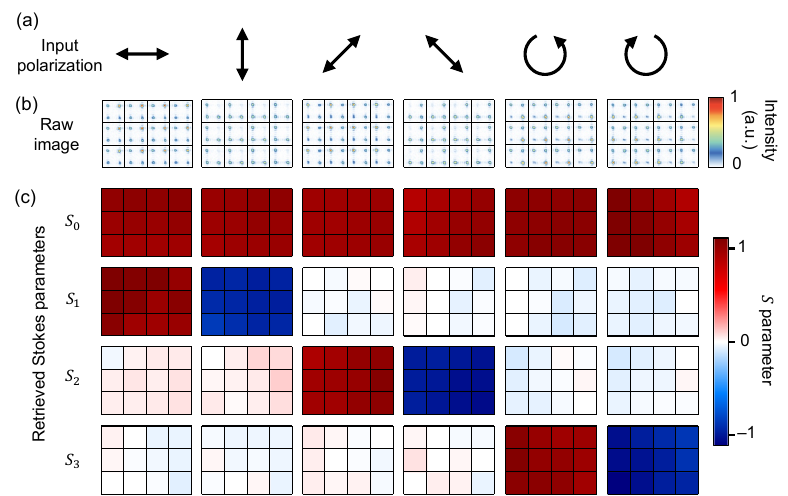}
\caption{Measured results of $3\times4$ superpixel array. (a) Input SOP. (b) Measured intensity distributions at the focal plane. (c) Retrieved Stokes parameters for each input SOP.}
\label{fig6}
\end{figure}

Figure~\ref{fig6} shows the results of evaluating the superpixel array by illuminating a plane-wave light onto the entire device.
Due to the limited size of the camera used for imaging, only $3\times4$ array out of $5\times5$ array is evaluated. 
Figure~\ref{fig6}(b) shows the measured intensity profile for six different input SOPs with $\vec{S}_\mathrm{in}=(1,\pm1,0,0)^t$, $(1,0,\pm1,0)^t$, and $(1,0,0,\pm1)^t$ as depicted in Fig.~\ref{fig6}(a). 
Using the integrated intensities at all pixels, 
the full Stokes parameters are retrieved by multiplying the matrix $\vb{B}'$
and displayed in Fig.~\ref{fig6}(c).
We can confirm that all 12 superpixels are operating correctly with an average detection error of $\ev{|\Delta\vec{S}_\mathrm{ret}|}=0.095$.

\begin{figure}[tb]
\centering\includegraphics{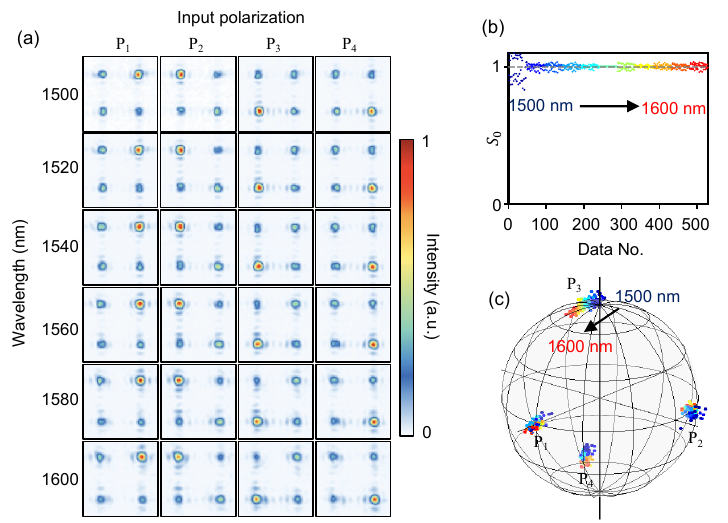}
\caption{Measured wavelength dependence. (a) Intensity distributions at a wavelength from 1500 nm to 1600 nm for the four different input SOPs P$_n$ ($n=1,2,3,4$). (b,c) Retrieved $S_0$ parameter and the Stokes vectors $(S_1, S_2, S_3)^t$ on the Poincaré sphere measured by all 12 superpixels at wavelength from 1500 nm to 1600 nm.}
\label{fig7}
\end{figure}

Finally, Fig.~\ref{fig7} shows the wavelength dependence of our metasurface measured from 1500 to 1600 nm.
Figure~\ref{fig7}(a) shows the measured intensity distributions at the focal plane of one superpixel, which indicates that our metasurface has a small wavelength dependence.
Figures~\ref{fig7}(b) and \ref{fig7}(c) show the retrieved $S_0$ and $(S_1, S_2, S_3)$ measured by all 12 superpixels over the entire wavelength range.
The average measurement errors $\ev{|\Delta\vec{S}_\mathrm{ret,\lambda}|}$ for all wavelengths are suppressed below 0.17, showing the broadband characteristics of our fabricated device.


\section{Conclusion}
\begin{table}[tb]
\centering
\caption{Comparison of different polarimeter configurations.}
\includegraphics{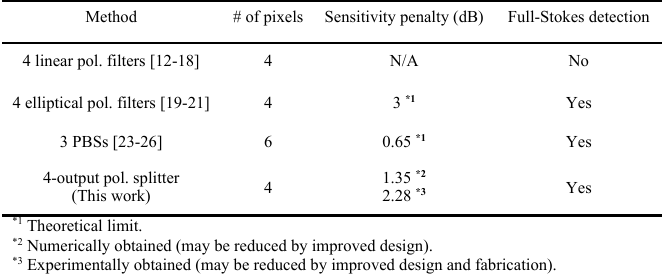}
\label{table1}
\end{table}
We have proposed and demonstrated a full-Stokes polarimeter using a single-layer dielectric metasurface with only four sensor pixels for high-sensitivity and high-resolution DoFP imaging polarimetry. 
Our metasurface serves as a four-output polarization splitter that projects input light onto four polarization states on a regular tetrahedron inscribed in the Poincaré sphere and then focuses them to different pixels at the focal plane. 
Although these four polarization states are not orthogonal to each other, we revealed numerically and experimentally that a single-layer metasurface enables low-loss splitting to achieve high imaging efficiency surpassing the 50\% theoretical limit of conventional four-pixel full-Stokes polarimeters that rely on polarization filtering.
Using a compact metasurface fabricated on an SOQ substrate, we achieved broadband full-Stokes detection across a wavelength range from 1500 to 1600 nm. Finally, simple imaging polarimetry was demonstrated using a $3\times4$ superpixel array. 

Table \ref{table1} compares this work with previously demonstrated polarimeters. Our polarimeter enables full-Stokes detection while keeping the number of sensor pixels to four, which is the minimum number required to detect full Stokes parameters.
Furthermore, the sensitivity penalties of our scheme are 1.35 dB and 2.28 dB from the numerical and experimental results, respectively, which are better than the 3-dB theoretical limit of the conventional polarization-filter-based full-Stokes polarimetry with four pixels.
While the size of the superpixel was $40\times40$ \textmu$\si{m^2}$ in this proof-of-concept demonstration, it could easily be scaled to accommodate higher-resolution image sensors with smaller pixel sizes. 
The demonstrated scheme can also be applied to other wavelength ranges, including the visible wavelength, by using appropriate meta-atom materials such as SiN$_\mathrm{x}$, GaN, and TiO$_2$.
With the low-loss non-orthogonal four-output polarization splitter that cannot be realized using bulk space optics, our metasurface-based full-Stokes polarimetry offers a unique and promising alternative of conventional polarization-filter-based polarization cameras, achieving higher sensitivity without sacrificing the spatial resolution.


\section*{Appendix}
\renewcommand{\thesection}{A\arabic{section}}
\renewcommand{\thefigure}{A\arabic{figure}}
\renewcommand{\theequation}{A\arabic{equation}}
\setcounter{section}{0}
\setcounter{figure}{0}
\setcounter{equation}{0}

\section{Sensitivity penalty $L$}
\label{sec:L}
In actual systems, the detected signals $\vec{I}$ contain various types of noise $\Delta \vec{I}$ including thermal noise and quantum noise. Then, the retrieved Stokes vector can be written as
\begin{equation}
    \vec{S}_\mathrm{ret} = \vec{S}_\mathrm{in} + \Delta \vec{S} = \vb{B}(\vec{I}+\Delta\vec{I}).
\end{equation}
$\Delta \vec{S}$ is the resultant error translated from $\Delta \vec{I}$ and is expressed as
\begin{equation}
    \Delta S_i = \sum_j b_{ij}\Delta I_j,
\end{equation}
where $\Delta S_i$ and $\Delta I_j$ denote the $i$-th and $j$-th components of $\Delta \vec{S}$ and $\Delta \vec{I}$. $b_{ij}$ is the $(i,j)$-th component of $\vb{B}$. 
To increase the sensitivity of polarimetry, a configuration with a smaller $|\Delta \vec{S}|$ is required. 
For simplicity, we assume that the noise at each port $\Delta I_j$ is independent and has the same standard deviation of $\sigma_I$. 
Then, the standard deviation of the error $\Delta S_i$ can be calculated as
\begin{equation}
    \sigma_{\Delta S_i} = \sqrt{\sum_j b_{ij}^2}\sigma_{I}.
\end{equation}
The sensitivity of a polarimeter can be evaluated quantitatively using a factor defined as \cite{Tanemura2020-nm}
\begin{equation}
    K \equiv \frac{\sqrt{\sum_i \sigma_{\Delta S_i}^2}}{\sigma_{I}}=\sqrt{\sum_{i,j} b_{ij}^2}.
    \label{eq:K}
\end{equation}
When $K$ is large, a larger error is incurred to the retrieved Stokes vector, degrading the sensitivity of the polarimeter.
In the following analysis, we assume that the total optical power incident on a single superpixel, $S_\mathrm{in0}$, is the same 
in all cases.
We also assume a perfect sensor responsivity and let $R=1$ for convenience.

First, we consider our proposed non-redundant configuration with an ideal lossless four-output polarization splitter and four sensor pixels as shown in Fig.~\ref{fig1}(c).
From Eq.~(\ref{eq:I}) with $\eta = 1$ and $R = 1$, the retrieval matrix $\vb{B}$ is written as
\begin{equation}
    \vb{B}=\vb{A}^{-1}=\qty[\frac{1}{4}\mqty(1&1&0&0\\1&-\frac{1}{3}&\frac{2\sqrt{2}}{3}&0\\1&-\frac{1}{3}&-\frac{\sqrt{2}}{3}&\frac{\sqrt{6}}{3}\\1&-\frac{1}{3}&-\frac{\sqrt{2}}{3}&-\frac{\sqrt{6}}{3})]^{-1}.
    \label{eq:B_our-case}
\end{equation}
From Eqs. (\ref{eq:K}) and (\ref{eq:B_our-case}), the $K$ factor in our case is derived as $K_\mathrm{ideal}=2\sqrt{10}$. 
For quantitative comparison, we define the penalty $L$ to describe the relative decrease in sensitivity as
\begin{equation}
    L \equiv \frac{K}{K_\mathrm{ideal}}.
\end{equation}


We now consider the case of the conventional polarization-filter-based full-Stokes polarimeter as shown in Fig.~\ref{fig1}(b). 
Similar to our case, the four SOPs of the polarization filters, P$_1$, P$_2$, P$_3$, and P$_4$, are chosen so that they constitute a regular tetrahedron on the Poincaré sphere as shown in Fig.~\ref{fig1}(e) to achieve highest sensitivity. 
Since the matrix $\vb{A}$ in this case is described by Eq.~(\ref{eq:I}) with $\eta =1/2$, we obtain $K_\mathrm{filter}=2K_\mathrm{ideal}=4\sqrt{10}$ and $L_\mathrm{filter}=2$. Note that this 3-dB sensitivity degradation comes from the transmission loss at the polarizers.

Finally, we consider the case of using three sets of PBS and six sensor pixels. 
In this case, the instrument matrix $\vb{A}$ is expressed as a $6\times4$ matrix,
\begin{equation}    
\vb{A}=\frac{1}{6}\mqty(1&1&0&0\\1&-1&0&0\\1&0&1&0\\1&0&-1&0\\1&0&0&1\\1&0&0&-1).
\end{equation}
Then, the $4\times 6$ retrieval matrix $\vb{B}$ is chosen to be the pseudo-inverse matrix of $\vb{A}$
\begin{equation}    
\vb{B}=\vb{A}^+=\mqty(1&1&1&1&1&1\\3&-3&0&0&0&0\\0&0&3&-3&0&0\\0&0&0&0&3&-3),
\end{equation}
to achieve the highest sensitivity.
The sensitivity factor and penalty are $K_\mathrm{PBS}=2\sqrt{15}$ and $L_\mathrm{PBS}=\sqrt{3/2}\sim0.65$ dB.

\begin{backmatter}
\bmsection{Funding}
This work was obtained in part from the commissioned research 03601 and 08801 by the National Institute of Information and Communications Technology (NICT), Japan, and was partially supported by Japan Society of Promotion of Science (JSPS) KAKENHI, Grant Number 24KJ0557.

\bmsection{Acknowledgments}
A part of this work was conducted at Takeda Sentanchi Supercleanroom, the University of Tokyo, supported by the “Nanotechnology Platform Program” of the Ministry of Education, Culture, Sports, Science and Technology (MEXT), Japan.
The SOQ wafer was provided by Shin-Etsu Chemical Co., Ltd. 

\bmsection{Disclosures}
The authors declare no conflicts of interest.

\bmsection{Data availability} Data underlying the results presented in this paper are not publicly available at this time but may be obtained from the authors upon reasonable request.

\bmsection{Supplemental document}
See Supplement~1 for supporting content. 

\end{backmatter}

\bibliography{references}

\end{document}